\documentstyle [12pt,a4p,epsfig,amsmath,multicol]{article}
\begin{document}
\vspace*{0.6cm}

\begin{center} 
{\normalsize\bf Lepton Flavour Eigenstates do not Exist if Neutrinos are Massive:
 `Neutrino Oscillations' Reconsidered}
\end{center}
\vspace*{0.6cm}
\centerline{\footnotesize J.H.Field}
\baselineskip=13pt
\centerline{\footnotesize\it D\'{e}partement de Physique Nucl\'{e}aire et 
 Corpusculaire, Universit\'{e} de Gen\`{e}ve}
\baselineskip=12pt
\centerline{\footnotesize\it 24, quai Ernest-Ansermet CH-1211Gen\`{e}ve 4. }
\centerline{\footnotesize E-mail: john.field@cern.ch}
\baselineskip=13pt
 
\vspace*{0.9cm}
\abstract{ If neutrinos are massive, `lepton flavour eigenstates' are absent
 from the amplitudes of all Standard Model processes. Measurements of 
 $\Gamma(\pi \rightarrow e \nu)/\Gamma(\pi \rightarrow \mu \nu)$ and of the
  MNS matrix are shown to exclude the existence of such states. Incoherent
 production of neutrinos of different flavour modifies the usual oscillation 
 phase derived assuming production of a lepton flavour eigenstate. The new
 prediction can be tested in long line-base experiments by comparison
 of `$\nu_{\mu}$ disappearance' of neutrinos from pion and kaon decay.}
\vspace*{0.9cm}
\normalsize\baselineskip=15pt
\setcounter{footnote}{0}
\renewcommand{\thefootnote}{\alph{footnote}}
\newline
PACS 03.65.Bz, 14.60.Pq, 14.60.Lm, 13.20.Cz 
\newline
{\it Keywords ;} Quantum Mechanics,
Neutrino Oscillations.
\newline

\vspace*{0.4cm}


  In the Standard Electroweak Model, the coupling of a charged lepton, $\ell_i$, of
  generation $i$ and a massive neutrino, $\nu_j$, of generation $j$  to the
   W-boson is described by
  the leptonic
 charged current:
 \begin{equation}
 J_{\mu}(CC)^{lept} = \sum_{i,j} \overline{\psi}_{\ell_i} \gamma_{\mu}(1-\gamma_5)U_{ij}\psi_{\nu_j}
 \end{equation}
  where $U_{ij}$ is the Maki-Nakagawa-Sakata (MNS)~\cite{MNS}
  charged lepton flavour/ neutrino mass
  mixing matrix. Table 1 shows the elements of this matrix as estimated~\cite{GGN}
 from experimental measurements of atmospheric and solar neutrino oscillations.
 The non-diagonal nature of this matrix gives evidence for strong violation
 of generation number (or lepton flavour) by $ J_{\mu}(CC)^{lept}$. Conservation of
 generation number corresponds to a diagonal MNS matrix with $\nu_1 = \nu_e$,
 $\nu_2 = \nu_{\mu}$ and  $\nu_3 = \nu_{\tau}$. This is the conventional massless
  neutrino scenario. With massive neutrinos and a non-diagonal MNS matrix, `$\nu_e$',
 `$\nu_{\mu}$' and `$\nu_{\tau}$' do not exist as physical states. That is, they do
  not appear in the amplitude for any physical process. This is because
  the neutrino charged 
 lepton couplings to the W are completely specified by  $J_{\mu}(CC)^{lept}$ which
 contains only the wavefunctions of the neutrino mass eigenstates $\nu_1$, $\nu_2$
  and $\nu_2$.

  \par Indeed, the introduction of such `lepton flavour eigenstates' as a linear
   superposition of neutrino mass eigenstates leads to predictions that are excluded
   by experiment. To show this, the pion decays: $\pi \rightarrow e \nu$ and
   $\pi \rightarrow \mu \nu$ may be considered. The invariant amplitude to produce
  the anti-neutrino mass eigenstate $\overline{\nu}_i$ in association with a charged lepton
  $\ell^-$ ($e^-$ or $\mu^-$) is:
  \begin{equation}
  {\cal M}_{\ell i} = \frac{G}{\sqrt{2}}f_{\pi}m_{\pi}V_{ud}
     \overline{\psi}_{\ell}(1-\gamma_5)U_{\ell i}\psi_{\overline{\nu}_i} 
  \end{equation}
  where $G$ is the Fermi constant, $f_{\pi}$ and $m_{\pi}$ are the 
  pion decay constant and mass respectively,  and
  $V_{ud}$ is an element of the CKM~\cite{CKM} quark flavour/mass mixing
  matrix. Introducing a `lepton flavour eigenstate': $\psi_{\overline{\nu}_\ell}$
  as a coherent superposition of the mass eigenstates  $\psi_{\overline{\nu}_i}$:
  \begin{equation}
   \psi_{\overline{\nu}_\ell} = U_{\ell 1}\psi_{\overline{\nu}_1}+
 U_{\ell 2}\psi_{\overline{\nu}_2}+U_{\ell 3}\psi_{\overline{\nu}_3}
 \end{equation}
 gives, for the invariant amplitude of the decay process $\pi^- \rightarrow 
 \ell^-\overline{\nu}_{\ell}$:
 \begin{equation}
  {\cal M}_{\ell} = {\cal M}^D_{\ell 1} U_{\ell 1} +  {\cal M}^D_{\ell 2} U_{\ell 2}
  + {\cal M}^D_{\ell 2} U_{\ell 3}
 \end{equation}
  where $ {\cal M}^D_{\ell i}$ ($D$ is for `diagonal') is given by Eqn(2) with the
  replacement $U_{\ell i}= 1$. If all neutrino masses are much smaller than the
  electron mass, it follows that:
 \begin{equation}
 {\cal M}^D_{\ell 1} \simeq  {\cal M}^D_{\ell 2} \simeq {\cal M}^D_{\ell 3}
  \simeq {\cal M}^D_{\ell 0}
 \end{equation}
 where ${\cal M}^D_{\ell 0}$ is the invariant amplitude calculated 
 on the assumption of massless neutrinos. With two flavour ($e$-$\mu$) mixing
 so that: 
 \begin{equation}
 \left(\begin{array}{cc}
    U_{e 1} & U_{e 2} \\
  U_{\mu 1} & U_{\mu 2}
 \end{array} \right) = 
  \left(\begin{array}{cc}
    \cos \theta_{12} &  \sin \theta_{12} \\
   - \sin \theta_{12} &  \cos \theta_{12}
 \end{array} \right) 
 \end{equation}
 and $U_{e 3} = U_{\mu 3} = 0$, Eqns(2)-(6) lead to the prediction:
 \begin{equation}
  R_{e/\mu} \equiv \frac{\Gamma(\pi^- \rightarrow e^- \overline{\nu}_e)}
 {\Gamma(\pi^- \rightarrow \mu^- \overline{\nu}_{\mu})} =
  \left(\frac{m_e}{m_{\mu}}\right)^2 \left [\frac{m_{\pi}^2-m_e^2}
   {m_{\pi}^2-m_{\mu}^2}\right]^2
  \left(\frac{\cos \theta_{12}+ \sin \theta_{12}}
   {\cos \theta_{12}- \sin \theta_{12}}\right)^2
  \end{equation}
  Allowing for radiative corrections~\cite{MS,GW} the world average experimental
  value $R_{e/\mu} = (1.230 \pm 0.004) \times 10^{-4}$ ~\cite{PDG}
  leads to a determination of the mixing angle $\theta_{12}$ via the relation:
  \begin{equation}
   \left(\frac{1+\tan \theta_{12}}{1-\tan \theta_{12}}\right)^2 = 0.9976\pm 0.0032
  \end{equation}
   The result is:
\[ \tan^2 \theta_{12} = (3.6^{+16.0}_{-3.2})\times 10^{-7} \]
 or \[  \tan^2 \theta_{12} < 3.0 \times 10^{-6} {\rm at~} 95 \% {\rm~CL} \]
   These results are evidently incompatible with the two-flavour mixing results
  obtained from the study of solar neutrino oscillations~\cite{GGN} where 
  $\theta_{\bigodot} \simeq \theta_{12}$:
  \[ 0.22 < \tan^2 \theta_{\bigodot} < 0.71~~ { \rm (LMA~ solution) } \]
  \[ 0.47 < \tan^2 \theta_{\bigodot} < 1.1~~ { \rm (LOW~ solution) } \]
   Allowing also for the coupling of the state $\nu_3$ to muons, as suggested
   by data on oscillations of atmospheric neutrinos, but still setting
    $U_{e 3} = 0$ ~\cite{KaysPDG} gives, in
   place of Eqn(8):
  \begin{equation}
   \left(\frac{\cos \theta_{12} +\sin \theta_{12}}
    {(\cos \theta_{12} -\sin \theta_{12})\cos \theta_{23}+
  \sin\theta_{23}}\right)^2
 = 0.9976\pm 0.0032
  \end{equation}

 \begin{table}
   \begin{center}
   \begin{tabular}{|c|c|c|c|} \hline  
       $j$  & 1 ($\nu_1$)  & 2 ($\nu_2$) &  3 ($\nu_3$) \\
   \hline \hline
      $i$  &   &   &    \\ \cline{1-1}
     1 ($e$)  & $0.79 \pm 0.12$   &  $0.57 \pm 0.16$ & $0.1 \pm 0.1$  \\
     2  ($\mu$) & $-0.45 \pm 0.25$  & $0.49 \pm 0.28$ & $0.69 \pm 0.18$  \\
    3  ($\tau$) & $0.34 \pm 0.29$    & $-0.60 \pm 0.23$   &  $0.67 \pm 0.18$ \\
  \hline
  \end{tabular}
   \caption[]{ Values of the MNS lepton flavour/neutrino mass mixing
   matrix $U_{ij}$ as derived from solar and atmospheric neutrino
   oscillation data~\cite{GGN}. }
  \end{center}
  \end{table} 

  Assuming $\sin \theta_{12} = 1/2$, and $\sin \theta_{23} = \cos \theta_{23} = 1/\sqrt{2}$
   as suggested in Reference~\cite{KaysPDG} (values consistent with the MNS matrix 
   elements shown in Table 1) gives the value 2 for the LHS of Eqn(9).
   It is clear, from these considerations, that the hypothesis
   that a coherent `lepton flavour eigenstate' is produced in pion decay is
   experimentally excluded, with an enormous statistical significance,
   by the experimental measurements of $R_{e/\mu}$ and the MNS matrix elements. 
   \par The correct prediction of the pion decay widths, that
   assumes incoherent production of the different neutrino mass eigenstates, is:
    \begin{eqnarray}
     \Gamma(\pi^- \rightarrow \ell^- \nu) & = & \Gamma(\pi^- \rightarrow \ell^- \nu_1)
      + \Gamma(\pi^- \rightarrow \ell^- \nu_2) +  \Gamma(\pi^- \rightarrow 
      \ell^- \nu_3)  \nonumber \\  
        & \simeq & {\cal M}_{\ell_0}^2 ( |U_{\ell 1}|^2 +|U_{\ell 2}|^2+ |U_{\ell 3}|^2)
        = {\cal M}_{\ell_0}^2
      \end{eqnarray}
 where the unitarity of the MNS matrix is invoked. The predicted value of $R_{e/\mu}$,
   given by setting $\theta_{12}=0$ in Eqn(7), is then 
  independent of the values of the elements of the MNS matrix. Unlike in the case
  where the coherent production of a  `lepton flavour eigenstate' is assumed,
  no information about the MNS matrix can be obtained from measurements of $R_{e/\mu}$
  \par Although the incoherent nature of the production of the different neutrino
  mass eigenstates, as exemplified in Eqn(10) above, was pointed out more than 20 years
  ago by Shrock~\cite{Shrock1,Shrock2}, and the unphysical nature of coherent states
  of neutrinos of different mass was also discussed in the literature~\cite{GKL1}
 the production of a coherent `lepton flavour eigenstate' at a fixed time remains 
  the basic assumption, in the literature, for the calculation of the phase of
  neutrino oscillations~\cite{KaysPDG}. The phase of each mass eigenstate is then assumed
   to evolve with time according to the Schr\"{o}dinger Equation in its own rest
   frame (the so-called `plane-wave' approximation). The assumption that all mass
   eigenstates are produced at the same time implicitly assumes equal velocities, since
   there is evidently a unique detection event at some well defined point in space-time.
   Still, in the derivation of the phase, the neutrino velocities, as defined by the
   kinematical relation: $v = p/E$ are assumed to be different. Thus contradictory 
   hypotheses are made in space-time and in momentum space. These assumptions lead to
  the so-called `standard formula' for the oscillation phase~\cite{KaysPDG}:
   \begin{equation}
    \phi_{ij}^{stand} = \frac{(m_i^2-m_j^2)}{2 p_{\nu}}L + O(m^4)
  \end{equation}
  where $m_i~(i=1,2,3)$ are the neutrino masses, $L$ is the distance between the 
     source pion and the detection event and $p_{\nu}$ is the neutrino momentum 
     in pion decay to a massless neutrino. Units with $\hbar = c = 1$ are used.
    \par It has been shown recently~\cite{JHF3} that the factor of two in the denominator
    of this equation is a consequence of a common 
    production time for all mass eigenstates, that follows from the incorrect 
    assumption that a coherent `lepton flavour eigenstate' is produced. As will 
     be demonstrated below, allowing for the possibility of different production
     times, due to the incoherent nature of the neutrino production process, doubles
     the contribution of neutrino propagation to the oscillation phase. 
   \par Other kinematical approximations that have been made in the literature, e.g.
     equal momenta and different energies or equal energies and different momenta, result
     in changes of only O($m^4$) to the phase, as compared to the exact result obtained
     by imposing both energy and momentum conservation in the neutrino
     production process. Thus all these results are
     equivalent at O($m^2$).

    \par In several previous papers written by the present author~\cite{JHF3,JHF,JHF2}
     results for the oscillation phase differing significantly from Eqn(11) have been
     obtained. These calculations were based on the covariant Feynman Path Integral 
     formulation of quantum mechanics. Here a concise re-derivation of the result
     for $e^-$ production resulting from the interaction of neutrinos from the
      decay  $\pi^+ \rightarrow \mu^+ \nu$ at rest will be given. For simplicity, the effect
       of the finite pion lifetime as well as that of the smearing of the physical masses of the
      $\pi$ and $\mu$ is neglected. As shown in Reference~\cite{JHF}, the corresponding
      corrections to the oscillation formula are completely negligible.
      The basis 
     of the calculation is very simple. Each interfering path amplitude has a common 
     initial state (a pion that comes to rest in a stopping target at laboratory time
      $t = t_0$) and a common final state (defined by the neutrino detection event). There
      is one such path amplitude for each neutrino mass eigenstate. As neither the neutrino's
      flavour nor its production time is observed, the corresponding amplitudes must be 
     coherently summed. Quantum mechanical coherence is a property of the
     path amplitudes (because they all have the same initial and final state) not of
     the neutrino production process. Indeed, the calculation is a direct application
     of the fundamental Eqn(7) of Feynman's original Path Integral paper~\cite{Feyn2}.
     \par As in the usual `plane-wave' derivation of Eqn(11)~\cite{KaysPDG},
        the space-time propagators of the neutrino
     and the source pion can be simply obtained by solving the time-dependent Schr\"{o}dinger
      Equation in the particle rest frame. The path amplitudes for the detection of an $e^-$
      produced in the quasi elastic scattering process: $\nu n \rightarrow e^- p$ 
      by a neutrino from the decay at rest: $\pi^+ \rightarrow \mu^+ \nu$ are then:
      \begin{eqnarray}
       A_{e \mu}(i) & = & A_0 U_{ei}\exp\{-im_i[\tau_D(\nu)-\tau_i(\nu)]\}U_{\mu i}
       \exp\{-im_{\pi}[\tau_i(\pi)-\tau_0(\pi)]\} \nonumber \\
         & = & A_0 U_{ei} U_{\mu i} \exp(-i\Delta \phi_i)~~~(i = 1,2,3)              
      \end{eqnarray}
       where:
       \begin{equation}
         A_0 \equiv \langle e^- p|T|\nu n\rangle \langle \nu \mu^+|T| \pi^+\rangle
       \end{equation} 
       The `reduced' transition amplitudes $\langle e^- p|T|\nu n\rangle$,
       $ \langle \nu \mu^+|T| \pi^+\rangle$ are defined by factorising out the MNS 
        matrix elements, e.g.:
         \begin{equation}
        \langle \nu_i \mu^+|T| \pi^+\rangle = U_{\mu i}\langle \nu \mu^+|T| \pi^+\rangle
         \end{equation}  
        In Eqn(12), $\tau_0(\pi)$,  $\tau_i(\nu,\pi)$ and $\tau_D(\nu)$ are the
        respective proper times of the $\pi$ or $\nu$ 
       at which the
       pion comes to rest in the stopping target, the neutrino is produced and the
       neutrino is detected.
       The overall propagator phase of the amplitude $A_{e \mu}(i)$ is the Lorentz invariant 
       quantity:
      \begin{equation}
        \Delta \phi_i = m_{i}[\tau_D(\nu)-\tau_i(\nu)]+m_{\pi}[\tau_i(\pi)-\tau_0(\pi)]    
       \end{equation}
       Since the pion is at rest:
     \begin{equation}
     \tau_i(\pi)-\tau_0(\pi) = t_i-t_0 = t_D -t_0 -\frac{L}{v_i} = t_D -t_0
       -L\left[1+\frac{m_i^2}{2 p_{\nu}}\right] + O(m^4)
       \end{equation}
       where $v_i$ is the velocity of $\nu_i$. 
     The exact kinematical relation for the neutrinos:
      \begin{equation}
       \Delta \tau = \frac{\Delta t}{\gamma} = \frac{m L}{E v} = \frac{m L}{p}
       \end{equation}
        and Eqn(16), substituted in Eqn(15), then give:
      \begin{equation}
        \Delta \phi_i = \frac{m_i^2}{p_{\nu}}\left[1-\frac{m_{\pi}}{2 p_{\nu}}\right]L+
        m_{\pi}(t_D-t_0-L)+ O(m^4)    
       \end{equation}   
        Eqns(12) and (18) then give, for the probability of $e^-$ detection        distance $L$        from the source:
         \begin{eqnarray} 
          P_{e \mu}(L) & = & \left| A_{e \mu}(1) + A_{e \mu}(2) + A_{e \mu}(3) \right|^2 
          \nonumber \\
             & = & -4 |A_0|^2 \left[ U_{e1}^{\ast} U_{\mu 1}^{\ast}U_{e2}U_{\mu 2} \sin^2
            \frac{\phi_{12}}{2}+ U_{e2}^{\ast} U_{\mu 2}^{\ast}U_{e3}U_{\mu 3} \sin^2
            \frac{\phi_{23}}{2} \right. \nonumber \\
          &  & \left .   + U_{e3}^{\ast} U_{\mu 3}^{\ast}U_{e1}U_{\mu 1} \sin^2
            \frac{\phi_{31}}{2}\right]             
          \end{eqnarray}   
        where 
     \begin{equation}
     \phi_{ij} = \frac{\Delta m^2_{ij}}{p_{\nu}}\left[1-\frac{m_{\pi}}{2 p_{\nu}}\right]L
       + O(m^4)     
     \end{equation}
        and 
      \begin{equation}
     \Delta m^2_{ij}  = m_i^2-m_j^2     
      \end{equation}

        \par For neutrinos produced in the decay of an arbitary particle
    at rest~\cite{JHF3,JHF}, or the decay, in flight, of an arbitary ultra-relativistic
    particle~\cite{JHF}, Eqn(20) generalises to:  
     \begin{equation}
     \phi_{ij} = \frac{\Delta m^2_{ij} L}{p_{\nu}}\frac{R_m^2}{(1-R_m^2)}
      + O(m^4)   
     \end{equation}
     where $R_m = m_X/m_S$ and $m_X$ is the mass of the particle (or
      system of particles) recoiling against the neutrino produced in the
     decay of a source particle of mass $m_S$. The neutrino is assumed to
    be light so that $m_S \gg m_i$. For $\beta$-decay at rest, the oscillation
    phase is given by~\cite{JHF3,JHF}:
     \begin{equation}
     \phi_{ij} = \frac{\Delta m^2_{ij}}{p_{\nu}}\left[1-\frac{E_{\beta}}{2 p_{\nu}}\right]L
       + O(m^4)
     \end{equation}
     where $E_{\beta}$ is the total energy release of the $\beta$-transition.
   \par The first term in the square brackets in Eqns(20) and (23) is the
    contribution to the oscillation phase of the neutrino propagator. It is 
    factor of two larger than the similar contribution given by the standard
    formula (11). The second term in the square brackets in Eqns(20) and (23)
    is the contribution to the oscillation phase from the propagator of the 
    source particle. It is opposite in sign to the neutrino contribution.
    It can be seen from Eqn(22), that, for light recoil masses such that
    $R_m \ll 1$, the oscillation phase tends to vanish, strong cancellation
    occuring between the neutrino and source particle contributions.
    \par Examination of Eqn(19) shows that the mechanism that governs the
     value of $P_{e \mu}$ is interference between the path amplitudes
     for different neutrino flavours. A small value of $P_{e \mu}$ is
     not an indication of an approximate conservation of lepton flavour,
     but of strong destructive interference between the different path
     amplitudes, independently of the values of the MNS matrix elements.
     This is made clear by consideration of the two flavour mixing scenario
     where $U_{e3} =0$ and Eqn(19) simplifies, using Eqn(6), to: 
     \begin{equation}
      P_{e \mu}(L) = |A_0|^2\sin^2 2 \theta_{12} \sin^2\frac{\phi_{12}}{2} =
      \frac{1}{2}|A_0|^2\sin^2 2 \theta_{12}(1-\cos\phi_{12}) 
     \end{equation} 
      The term $-\cos\phi_{12}$ in Eqn(24) originates in the interference of the
     path amplitudes corresponding to $\nu_1$ and $\nu_2$. For small values of
     $L$, $e^-$ production is suppressed by the almost complete destructive 
     interference of these amplitudes, independently of the value of $\theta_{12}$
     i.e. of the degree of non-conservation of lepton number. The destructive
     nature of the interference is due to the minus sign multiplying
     $\sin \theta_{12}$ in the second row of the matrix on the RHS of Eqn(6).
     This, in turn, is a consequence of the unitarity of the MNS matrix.  
     \par Indeed, nowhere in the description of the `$\nu_e$ appearence' experiment, 
     just presented, do `lepton flavour eigenstates' occur, although such
    an experiment is typically referred to~\cite{KaysPDG} as
     `$\nu_{\mu} \rightarrow \nu_e$ 
     flavour oscillation'. In fact, only the mass eigenstates $\nu_1$, $\nu_1$
     and $\nu_3$ appear in the amplitudes of the physical
     processes which interfere. It is the interference of these amplitudes
     to produce the detection event that constitutes the phenomenon of
     `neutrino oscillations'. Indeed, no temporal oscillations of
      `lepton flavour' actually occur. Still the terms `$\nu_e$',
       `$\nu_{\mu}$' and  `$\nu_{\tau}$' may still have a certain
       utility as mnemonics, even though they do not represent
       physical neutrino states for massive neutrinos. For example, it makes
        sense to 
       refer to solar neutrinos, in a loose way, as a  `$\nu_e$ beam '
       since the different physical components are all created together
      with an electron. Similarly, atmospheric neutrinos are 
     predominantly  `$\nu_{\mu}$', i.e., born together with a muon. 
     \par The values of the correction factor $C$ that must be applied to 
     a value of $\Delta m_{ij}^2$ obtained using the standard formula (11), to
     obtain the corresponding quantity as derived from the path amplitude
     formulae of Eqns(20), (22) or (23) are presented in Table 2 for 
     different neutrino sources.
      The correction factor is:
     \begin{equation}
      C = \frac{\phi_{ij}^{stand}}{\phi_{ij}} = \frac{(\Delta
       m_{ij}^2)^{stand}}{\Delta m_{ij}^2} = 1/(\frac{E_S}{p_{\nu}}-2)
       = \frac{1-R_m^2}{2 R_m^2}
      \end{equation}
       where $E_S = m_S$ for particle decays and  $E_S = E_{\beta}$ for
       $\beta$-decays.

   \begin{table}
   \begin{center}
   \begin{tabular}{|c|c|c|} \hline  
     Process   & Kinematical Condition   & $C$  \\
   \hline \hline
    $\mu \rightarrow e \nu \nu$ &   &    \\
  $\tau \rightarrow e \nu \nu$ &   $\langle p_{\nu} \rangle = 7 m_{\ell}/20$   &  1.16 \\
    $\tau \rightarrow \mu \nu \nu$ &   &    \\
 \hline
   $\pi \rightarrow \mu \nu$ & - & 0.372 \\
 \hline
   $\pi \rightarrow e \nu$ & - & 37300 \\
 \hline
   $K \rightarrow \mu \nu$ & - & 10.4 \\
 \hline
   $K \rightarrow e \nu$ & - & 466718 \\
 \hline
   $K \rightarrow \pi \mu \nu$ & $m_X = m_{\pi}+m_{\mu}$  ($C^{max}$) & 1.52 \\
 \hline
    $K \rightarrow \pi e \nu$ & $m_X = m_{\pi}+m_{e}$  ($C^{max}$) & 5.71 \\
 \hline
     &   &  \\ 
  $^Z N_A \rightarrow ^{Z-1} N_A+ e^+ +\nu$ & $p_{\nu}/ p_{\nu}^{max} = 2/3$
  & 1.0  \\
  & e.g. $E_{\nu} = 9.3$ MeV & \\
  & for $^{8}B \rightarrow ^{8}Be^{\ast}+ e^++ \nu$ &  \\
 \hline 
 $ e^{-}+ ^{Z}N_A \rightarrow ^{Z-1}N_A+ \nu$ & e.g.  $ e^{-}+ ^{7}Be \rightarrow ^{7}Li+ \nu$ & 1.0 \\
 \hline
  \end{tabular}
   \caption[]{ Values of the correction factor $C$ relating the oscillation
    phase, as calculated using the standard formula, to that calculated using
    path amplitudes for various neutrino sources. }
  \end{center}
  \end{table} 
   
  \par For muon decays and purely leptonic $\tau$ decays, at the mean decay
   neutrino energy, as in the first row of Table 2, the correction to
    $(\Delta m_{ij}^2)^{stand}$ is only 16 $\%$. The small recoil masses in the
    decays: $\pi \rightarrow e \nu$, ${\rm K} \rightarrow e \nu$ lead to very
    large values of $C$ (i.e. small values of the oscillation phase). For the
     electron capture process in the last row of Table 2, which is important
    for solar neutrino production, the correction factor is unity. The
    same is true in $\beta$-decay processes when $p_{\nu}/ p_{\nu}^{max} = 2/3$,
    close to the mean value  $\langle p_{\nu} \rangle = (11/16) p_{\nu}^{max}$ of the
    spectrum of allowed $\beta$-decays. There is then hope that mass differences
    derived from reactor neutrino experiments~\cite{KAMLAND} using the formula (11)
     will not be too different from those given by the path amplitude formula
     (23). Actually, the phase should be evaluated separately for each $\beta$-decay
    process contributing to the reactor flux, and a suitable weighted average 
    performed. This is evidently a complex and difficult undertaking. In contrast
    the corrections for atmospheric neutrinos (predominantly from
      $\pi \rightarrow \mu \nu$ decay) and high energy solar neutrinos, essentially
     uniquely from $^{8}B$ $\beta$-decay (as shown in the last-but-one row of Table 2)
     can be applied in a straightforward manner.
    \par Direct experimental discrimination between the standard formula for the
     oscillation phase (11) and the path amplitude formula (22) is possible in 
     long-baseline terrestrial oscillation experiments by comparing `$\nu_{\mu}$
     disappearence' as observed in neutrino beams produced by either 
     $\pi \rightarrow \mu \nu$ or  ${\rm K} \rightarrow \mu \nu$  decays.
     Because of the relatively smaller recoil mass in the second process
     there is a strong relative suppression of the interference phase. For fixed 
     neutrino momentum and source-detector distance, and considering only
     `$\nu_{\mu} \rightarrow \nu_{\tau}$ oscillations':
     \begin{equation}
      \phi_{23}({\rm K}) = \frac{C_{\pi}}{C_{{\rm K}}}  \phi_{23}(\pi) = 0.0358
      \phi_{23}(\pi)  
     \end{equation}
     The K2K experiment with a 250 km baseline has recently found indications
     of `$\nu_{\mu}$ disappearence' for neutrinos from the
    process  $\pi \rightarrow \mu \nu$ of average energy 1.3 GeV~\cite{K2K}.
    Using the standard formula (11)
    the event best fit to the data 
    gives $\Delta m_{23}^2 = 2.8 \times 10^{-3}$ eV$^2$ and
     $\theta_{23} \simeq \pi/4$~\cite{K2K}. Inserting $L = 250$km in Eqn(11)
      gives a phase $\phi_{23} = 1.37$ rad.
      Inserting these values into the two flavour `$\nu_{\mu}$ disappearence'
     formula~\cite{KaysPDG}:
     \begin{equation}
      P_{\mu \mu}(L) \simeq 1-\sin^22\theta_{23}\sin^2\frac{\phi_{23}}{2}
     \end{equation}
     gives a suppression factor of 0.6 as compared to the no-oscillation
     ($\phi_{23}=0$) case. Formula (11)
     predicts exactly the same phase and suppression
      factor for neutrinos of the same average momentum
    produced in  ${\rm K} \rightarrow \mu \nu$ decays. However, if Eqns(22) 
    and (26) are correct,  $\phi_{23}(K) = 49$ mrad and a reduction of
    only 0.06 $\%$ of the `$\nu_{\mu}$ flux' is predicted. This crucial 
    test of the basic quantum mechanics of neutrino oscillations requires 
    the installation of separated low energy charged kaon beams in 
    the JHF-SK complex~\cite{JHFSK}, or in similar long-linebase neutrino
    facilities.

\pagebreak

\end{document}